\documentclass[pra,showpacs,preprintnumbers,amsmath,amssymb]{revtex4}
\usepackage{bm}
\usepackage{amsmath}
\usepackage[dvips]{graphicx}
\begin{document}
\bibliographystyle{apsrev}

\title{ Finite nuclear size effect on Lamb shift of
s$_{1/2}$, p$_{1/2}$, and p$_{3/2}$ atomic states}

\author{A. I. Milstein}
\email[Email:]{A.I.Milstein@inp.nsk.su} \affiliation{Budker
Institute of Nuclear Physics, 630090 Novosibirsk, Russia}
\author{O. P. Sushkov}
\email[Email:]{sushkov@phys.unsw.edu.au} \affiliation{School of
Physics, University of New South Wales, Sydney 2052, Australia}
\author{I. S. Terekhov}
\email[Email:]{I.S.Terekhov@inp.nsk.su} \affiliation{ Novosibirsk
University, 630090 Novosibirsk, Russia}

\date{\today}
\begin{abstract}
We consider one-loop self-energy and vacuum polarization radiative
corrections to the shift of atomic energy level due to finite
nuclear size. Analytic expressions for vacuum polarization
corrections are derived. For the self-energy of $p_{1/2}$ and
$p_{3/2}$ states in addition to already known terms we derive
next-to-leading nonlogarithmic $Z\alpha$-terms. Together with
contributions obtained earlier the terms derived in the present
work give explicit analytic expressions for $s_{1/2}$ and
$p_{1/2}$ corrections which agree with results of previous
numerical calculations  up to Z=100 (Z is the nuclear charge
number). We also show that the finite nuclear size radiative
correction for a $p_{3/2}$ state is not small compared to the
similar  correction for a $p_{1/2}$ state at least for small $Z$.

\end{abstract}
\pacs{11.30.Er, 31.30.Jv, 31.30.Gs}  \maketitle

\section{Introduction}\label{I}

Experimental and theoretical investigations of the radiative shift
(Lamb shift) of energy levels in heavy atoms and ions is an
important way to test Quantum Electrodynamics in the presence of a
strong external electric field. One of the effects related to this
problem is the dependence of the Lamb shift on the finite nuclear
size (FNS). One can also look at this effect from another point of
view. It is well known that there is an isotope shift of atomic
levels due to the FNS. Corrections we are talking about are the
radiative corrections to the isotope shift. There are two types of
corrections. The first type is due to vacuum polarization and we
will call them VPFNS corrections. The second type is due to the
electron self-energy and vertex and we will call them SEVFNS
corrections. Calculations of both VPFNS and SEVFNS corrections
have attracted a lot of attention, for a review see Ref.
\cite{MPS}. Until recently it was mainly numerical work. The
corrections for $1s$, $2s$, and $2p$ states have been calculated
numerically, exactly in $Z\alpha$, in
Refs.\cite{M,JS,CJS93,Mohr,Blun,LPSY,BMPS};  see also review
\cite{Borie} for muonic atoms . Here $Z$ is the nuclear charge
number and $\alpha$ is the fine structure constant. Analytical
calculations have been  based on $Z\alpha$-expansion. The s-wave
VPFNS correction has been calculated a long time ago in order
$\alpha(Z\alpha)$ \cite{F,H}. An ultraviolet logarithmic
enhancement of higher order contributions to the VPFNS correction
has been revealed in Ref. \cite{Mil0}. The enhancement factor is $
\ln^2(\lambda_C/r_0)$, where $\lambda_C$ is the Compton wave
length and $r_0$  is the nuclear radius. In Ref. \cite{Mil0} the
double-logarithmic contribution to the radiative correction to
atomic parity nonconservation has been calculated analytically
exactly in $Z\alpha$. The formula derived in \cite{Mil0} gives
also one of the double-logarithmic contributions to  the VPFNS
correction for $s_{1/2}$ and $p_{1/2}$ states. The SEVFNS
correction for an $s$-wave state has been calculated analytically
in order $\alpha(Z\alpha)$ in Refs.\cite{Pach93,PG,Mil,Mil1}, see
also reviews \cite{rev1,rev2}. Structure of higher-order
contributions to the SEVFNS correction has been elucidated in our
recent papers \cite{Mil,Mil1}. The contributions are also
logarithmically enhanced, but the enhancement factor contains only
first power of $ \ln(\lambda_C/r_0)$. For $s_{1/2}$ and $p_{1/2}$
states prefactor before the logarithm has been calculated in Ref.
\cite{Mil} in order $\alpha(Z\alpha)^2$ and in Ref. \cite{Mil1}
exactly in $Z\alpha$. Finally, contributions of the order
$\alpha(Z\alpha)^0$ to SEVFNS corrections for $p_{1/2}$ and
$p_{3/2}$ states have been calculated in recent papers
\cite{Mil2,Jen}.

In the present work we first  calculate
nonlogarithmic $\alpha(Z\alpha)$ terms of SEVFNS
corrections for $p_{1/2}$ and $p_{3/2}$ states. Together with
previously known contributions this gives a full analytic
description for $p_{1/2}$ SEVFNS correction which agrees with
previous numerical results up to $Z=100$. Our
calculation demonstrates also that the $p_{3/2}$ SEVFNS correction
is comparable with the $p_{1/2}$ SEVFNS correction. The second part of
the work is devoted to analytical calculations of VPFNS
corrections for $s_{1/2}$ and $p_{1/2}$ states. Analytic formulas
obtained in this paper agree with previous numerical results up to
$Z=100$.

The structure of the present paper is as follows. In Sec. II
the general structure of SEVFNS and VPFNS corrections is
discussed. In Sec. III we calculate $\alpha(Z\alpha)$ terms for
$p_{1/2}$ and $p_{3/2}$ SEVFNS corrections, summarize all known
terms for s-wave and p-wave corrections and compare analytic
expressions with previous numerical results.In Sec. IV
we first calculate unknown contributions to the VPFNS correction. Then
we summarize contributions calculated in the present work and corrections
obtained before and compare our analytic results with previous numerical
calculations. Finally, Sec. V presents our conclusions.

\section{General structure of SEVFNS and VPFNS radiative corrections}
\label{II}

\begin{figure}[ht]
\centering
\includegraphics[height=200pt,keepaspectratio=true]{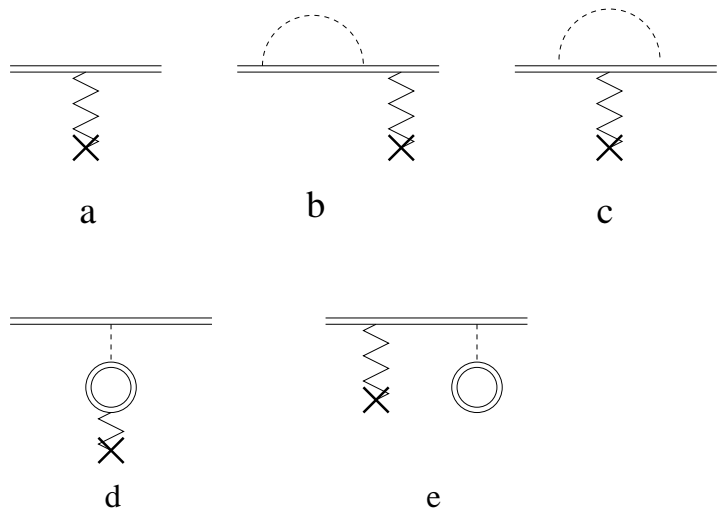}
\caption{\it The leading contribution to the FNS
effect is given  by diagram (a), and one loop radiative
corrections to the effect are given by diagrams (b-e). The double
line denotes exact electron Green's function in the Coulomb
field of the nucleus, the cross denotes the nucleus, the zigzag
line denotes the FNS perturbation (\ref{dV}), and the dashed line
denotes the photon. } \label{Fig1}
\end{figure}

Due to the finite nuclear size, the electric potential $V(r)$ of
the nucleus differs  from that for a point-like
nucleus. The deviation is
\begin{equation}
\label{dV}
\delta V(r)=V(r)-\left(-\frac{Z\alpha}{r}\right)
\end{equation}
Throughout the paper we set $\hbar=c=1$. The diagram that describe
the FNS effect in the  leading order is shown in
Fig.\ref{Fig1}a.
\noindent The double line corresponds to the exact electron wave
function in the Coulomb field, the zigzag line with cross denotes
the perturbation (\ref{dV}), and the dashed line denotes the
photon. In momentum representation the perturbation (\ref{dV})
reads
\begin{eqnarray}
\label{dV1} \delta V(k)= \frac{4\pi Z\alpha}{k^2}
 [1-{\cal F}(k^2)]\, ,
\end{eqnarray}
where ${\cal F}(k^2)$  is the nuclear form factor. An important
fact is that (\ref{dV}) is a local interaction which is nonzero
only at $r \sim r_0$, where $r_0$ is the nuclear radius. For
estimates one can use the following formula for $r_0$
\begin{equation}
\label{r0} r_0 \approx 1.1 \ A^{1/3}\ fm \approx 1.5 \ Z^{1/3} \
fm \ ,
\end{equation}
where $A$ is the mass number of the nucleus. Dynamics of electrons
at $r\sim r_0$ is ultrarelativistic at any value of  $Z$. The
Dirac wave function at $r\ll Z\alpha \lambda_C$ is of the form,
see Ref. \cite{BLP}
\begin{equation}
\label{Dirac} \Psi({\bm r})= Nr^{\gamma-1}\begin{pmatrix}
(\kappa-\gamma)\Omega\\
iZ\alpha\tilde{\Omega}
\end{pmatrix}
\quad ,
\end{equation}
where $\Omega$ and $\tilde{\Omega}=-({\bm \sigma}\cdot{\bm
n})\Omega$ are spherical spinors; $\kappa=-1$ for $s_{1/2}$ state,
$\kappa=1$ for $p_{1/2}$ state, and $\kappa=-2$ for $p_{3/2}$
state; $\gamma=\sqrt{\kappa^2-(Z\alpha)^2}$; and $N$ is a constant
known for each particular state. For hydrogenlike ions one can
find $N$ using wave functions presented in Ref \cite{BLP}. For
multi-electron ions or atoms the constant $N$ can be calculated
numerically. It can also be calculated analytically but in the
semiclassical approximation \cite{Kh}.

Due to ultrarelativistic nature of the problem the strong
relativistic enhancement is a special
property of SEVFNS and VPFNS radiative corrections. This is similar to atomic
parity nonconservation (PNC) \cite{Kh}. The relativistic
enhancement factor for $s_{1/2}$ and $p_{1/2}$ states is
proportional to $R_{rel} \sim (\lambda_C/Z\alpha
r_0)^{2(1-\gamma)}$, The factor is $R_{rel} \approx$3 for Cs and
$R_{rel} \approx 9$ for Tl, Pb, and Bi \cite{Kh}. The enhancement
factor $R_{rel}$ is divergent at $r_0 \to 0$.  As a
result behavior of SEVFNS, VPFNS, and PNC radiative
corrections  differs essentially from that  of radiative
corrections to the hyperfine structure.

The FNS energy shift $\Delta E$ in the leading approximation,
diagram Fig.1a, has been calculated in numerous works both
analytically and numerically. For a hydrogenlike ion there are
simple parameterizations suggested in Ref. \cite{Shab}
\begin{eqnarray}
\label{Sh}
\Delta E_{ns_{1/2}}&=&\frac{Z^2}{10n}[1+(Z\alpha)^2f_{ns_{1/2}}]
\left(\frac{2Z\alpha R}{n\lambda_C}\right)^{2\gamma} \ ,\nonumber\\
\Delta E_{np_{1/2}}&=&\frac{Z^4\alpha^2(n^2-1)}{40n^3}[1+
(Z\alpha)^2f_{np_{1/2}}]
\left(\frac{2Z\alpha R}{n\lambda_C}\right)^{2\gamma} .
\end{eqnarray}
The energies are given in atomic units,  $E_0=m\alpha^2=27.2eV$;
 $m$ is  the electron mass;
$n$ is the principal quantum number, $n=1,2,...$;
$R\approx r_0$ is some effective radius (for details see
Ref. \cite{Shab}); and coefficients $f_i=f_i(Z\alpha)$ are
\begin{eqnarray}
\label{fi}
f_{1s_{1/2}}(Z\alpha)&=&1.380-0.162(Z\alpha)+1.612(Z\alpha)^2 \ ,
\nonumber \\
f_{2s_{1/2}}(Z\alpha)&=&1.508+0.215(Z\alpha)+1.332(Z\alpha)^2 \ ,
\nonumber \\
f_{2p_{1/2}}(Z\alpha)&=&1.615+4.319(Z\alpha)-9.152(Z\alpha)^2 +
11.87(Z\alpha)^3 \ .
\end{eqnarray}
To be absolutely correct we should say that
expectation values of the perturbation (\ref{dV}) taken with the
wave functions (\ref{Dirac}) are not sufficient to calculate the
energy shifts. One has also to take into account deformation of
the wave function inside the nucleus. However, Eqs. (\ref{Sh})
take this effect into account, and as soon as it is done there is
no need to care about this detail any more.

Diagrams Fig.\ref{Fig1}(b-e) show the radiative
corrections under discussion, see also comment \cite{com0}. Diagrams
Fig.\ref{Fig1}b (self-energy) and Fig.\ref{Fig1}c (vertex) describe the
SEVFNS correction and diagrams Fig.\ref{Fig1}d
and Fig.\ref{Fig1}e describe the VPFNS correction.
Note that  Fig.\ref{Fig1}d corresponds to a
modification of $\delta V$ (see Eq. (\ref{dV})) due to the vacuum
polarization, and Fig.\ref{Fig1}e corresponds to a
modification of the electron wave function due to the vacuum polarization.
Each of the diagrams Fig.\ref{Fig1}(b-e) gives some energy shift
$\delta E_i$, i=b,c,d,e. Let us stress that for $s_{1/2}$ and $p_{1/2}$
states we {\it always} calculate a
relative value of the radiative correction
\begin{eqnarray}
\label{ddef}
\Delta_{s_{1/2}}&=&\delta E_{s_{1/2}}/\Delta E_{s_{1/2}} \ , \nonumber\\
\Delta_{p_{1/2}}&=&\delta E_{p_{1/2}}/\Delta E_{p_{1/2}} \ ,
\end{eqnarray}
where $\Delta E$ is the FNS shift given by diagram Fig.\ref{Fig1}a
for the {\it same} atomic state. To find the value of $\Delta E$
one can use Eqs. (\ref{Sh}), or a numerical calculation or any
other method. Situation with $p_{3/2}$ correction is different and
we discuss this at the end of section III (see Eq. (\ref{ddef1}).
So, a crucially important point is that we calculate  analytically
{\it only} the relative quantity $\Delta$  . This is the central
issue of this work as well as of our recent papers
\cite{Mil0,Mil,Mil1,Mil2}. The point is that convergence of
$Z\alpha$-expansion for $\Delta$ is quite reasonable while the
ultrarelativistic  character of the FNS effect mentioned above
makes the convergence of $Z\alpha$-expansion for $\delta E$
 very poor. The good convergence of the series for the
quantity $\Delta$ is related to separation of scales. Radiative
corrections which do not contain $\ln(\lambda_C/r_0)$ are related
to quantum fluctuations at "large" distances, $r\sim \lambda_C$.
On the other hand, the perturbation is nonzero only at small
distances $r \sim r_0 \ll \lambda_C$. Therefore, the
 enhancement related to small distances in the
contribution of "large distance" quantum fluctuations  is
factorized similar to (\ref{Sh}) , in essence this is a kind of
effective operator approach. There are also quantum fluctuations
from distances $r_0 < r < \lambda_C$, their contribution cannot be
trivially factorized and we take special care about these
fluctuations. This is where terms dependent on
$\ln(\lambda_C/r_0)$ come from.  The factorization of small
distance enhancement in the part of the FNS effect related to the
vacuum polarization was implicitly demonstrated in Ref.
\cite{Lee}.

\section{Self-energy and vertex FNS radiative corrections}\label{III}
\begin{figure}[ht]
\centering
\includegraphics[height=180pt,keepaspectratio=true]{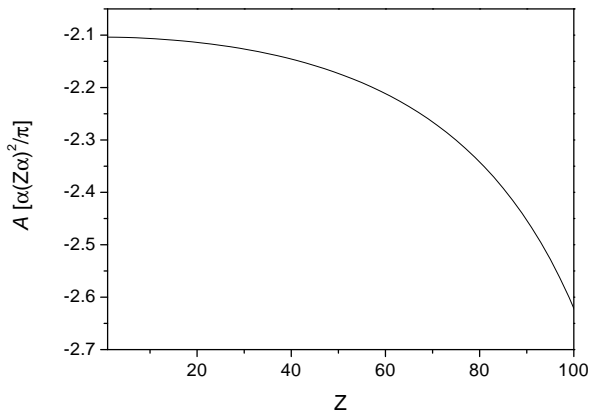}
\caption{\it The function ${\cal A}$ , Eq.(\ref{general}),
calculated to all orders in $Z\alpha$. Value of the function is
given in units $\alpha(Z\alpha)^2/\pi$. } \label{Fig2}
\vspace{10pt}
\end{figure}

\begin{figure}[ht]
\centering
\includegraphics[height=200pt,keepaspectratio=true]{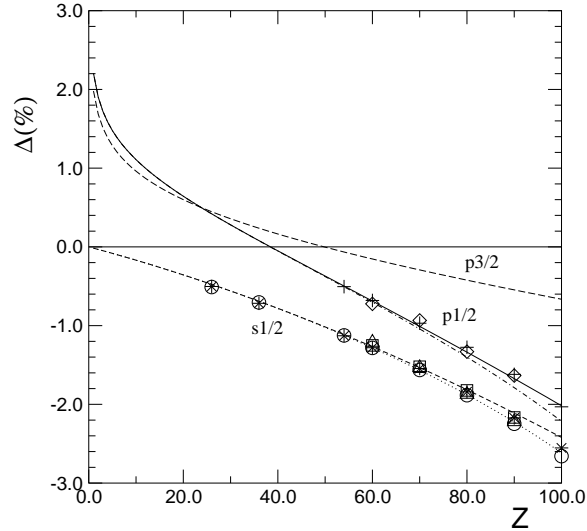}
\vspace{10pt} \caption{\it Relative SEVFNS corrections (\%) for
$s_{1/2}$ and $p_{1/2}$ states. The dashed line shows the
correction $\Delta_{s_{1/2}}^{SEV}$ given by Eq. (\ref{ssev}). The
same correction, but with $\mathcal{A}$ taken from Fig.\ref{Fig2}
is  plotted by the dotted line. Results of computations of
$\Delta_{s_{1/2}}^{SEV}$ for $1s$ states are shown by squares
\cite{CJS93} and stars \cite{Mohr}. Similar results for $2s$
states are shown by triangles \cite{CJS93} and circles
\cite{Mohr}. The solid line shows the correction
$\Delta_{p_{1/2}}^{SEV}$ given by Eqs.
(\ref{general}),(\ref{AA}),(\ref{B1}). The same correction, but
with $\mathcal{A}$ taken from Fig.\ref{Fig2} is plotted by the
dashed-dotted line. Results of computations of
$\Delta_{p_{1/2}}^{SEV}$ for $2p_{1/2}$ states are shown by
diamonds \cite{CJS93} and crosses \cite{Mohr}. The correction
$\Delta_{p_{3/2}}^{SEV}$ given by Eq. (\ref{B13}) is plotted  by
the long-dashed line. } \label{Fig3}
\end{figure}

Technically the most complicated  are the self-energy and the
vertex  FNS  radiative  corrections (SEVFNS) given by diagrams in Fig.\ref{Fig1}b
and Fig.\ref{Fig1}c.
It has been demonstrated in paper \cite{Mil0} that relative SEVFNS corrections
for $s_{1/2}$ and $p_{1/2}$ states are of the form
\begin{equation}
\label{general}
\Delta^{SEV} = \mathcal{A}\ln(b\lambda_C/r_0) + \mathcal{B}\, ,
\end{equation}
where $\mathcal{A}$ and $\mathcal{B}$ are functions of
$Z\alpha$ independent of $r_0$; $b=\exp(1/(2\gamma)-C-5/6)$, $\gamma =
\sqrt{1-(Z\alpha)^2}$, $C\approx 0.557$ is the Euler constant.
The function $\mathcal{A}(Z\alpha)$ is the same for $s_{1/2}$ and $p_{1/2}$
states. This function has been calculated in the leading order in $Z\alpha$ in
Ref. \cite{Mil}, the result reads
\begin{equation}
\label{AA}
\mathcal{A}=
-\frac{\alpha(Z\alpha)^2}{\pi}\left(\frac{15}{4}-\frac{\pi^2}{6}\right) \ .
\end{equation}
It has been also calculated {\it exactly} in $Z\alpha$ in Ref. \cite{Mil1}.
Plot of $\mathcal{A}$ from Ref. \cite{Mil1} is presented in Fig.\ref{Fig2}.
Deviation of (\ref{AA}) from exact $\mathcal{A}(Z\alpha)$
is small even at $Z=100$.
The function $\mathcal{B}(Z\alpha)$ for an $s_{1/2}$ state has been calculated
in the leading order in $Z\alpha$ in Refs. \cite{Mil,Mil1}
\begin{equation}
\label{BB}
\mathcal{B}_{s}=
-\alpha (Z \alpha)\left(\frac{23}{4}- 4\ln 2\right) \ .
\end{equation}
It agrees numerically with earlier results \cite{Pach93,PG}.
Combining Eqs. (\ref{general}),(\ref{AA}),(\ref{BB}) one gets the
the following expression for s-wave SEVFNS correction
\begin{equation}
\label{ssev}
\Delta^{SEV}_{s_{1/2}} = -\alpha (Z \alpha)\left(\frac{23}{4}- 4\ln 2\right)
-\frac{\alpha(Z\alpha)^2}{\pi}\left(\frac{15}{4}-\frac{\pi^2}{6}\right)
\ln(b\lambda_C/r_0) \ .
\end{equation}
The leading unaccounted term is of the order $\sim
\alpha(Z\alpha)^2/\pi$. The correction (\ref{ssev}) in \%  is
plotted in Fig.\ref{Fig3} by the dashed line. The same correction,
but with $\mathcal{A}$ taken from Fig.\ref{Fig2} is plotted in
Fig.\ref{Fig3} by the dotted line.
Our analytical results for $\Delta_{s_{1/2}}^{SEV}$ are in a good
agreement with previous numerical data  \cite{CJS93,Mohr} shown by
squares and stars for 1s state and by triangles and circles for 2s
state. Some dependence of $\Delta_{s_{1/2}}^{SEV}$ on the
principle quantum number $n$ can appear from function
$\mathcal{B}_{s}$ in the order $\alpha(Z\alpha)^2/\pi$. However,
one can see from the numerical data \cite{CJS93,Mohr} presented in
Fig. \ref{Fig3} that this dependence remains very weak up to
$Z=100$.

The function $\mathcal{B}_{p1/2}$ for $p_{1/2}$ states in the leading order has
been calculated in Refs. \cite{Mil2,Jen},
\begin{equation}
\label{B0}
\mathcal{B}_{p1/2}^{(0)}=\alpha\frac{8}{9\pi}\left(\ln\frac{1}{(Z\alpha)^2}
+ d\right) \ .
\end{equation}
For $2p_{1/2}$ state the constant is $d=0.910$. This constant
 depends slightly on the principle quantum number $n$, see Ref.
\cite{Jen}.

It is interesting to note that $Z\alpha$-expansion of
$\mathcal{B}_{p1/2}$ starts from $\alpha\ln(Z\alpha)$ term, while
the same expansion for $\mathcal{B}_{s}$ starts from $\alpha(Z
\alpha)$. It is related to a different infrared behavior of
quantum fluctuations, see discussion in Ref. \cite{Mil2}.

Now we calculate the $\alpha(Z\alpha)$ term in the expansion
of $\mathcal{B}_{p1/2}$. The electron  wave function at small
distances is given by Eq. (\ref{Dirac}). Let us first calculate
the leading contribution to the FNS shift given by diagram in
Fig.\ref{Fig1}a. For $s_{1/2}$ and $p_{3/2}$ states the upper
component of the Dirac spinor (\ref{Dirac}) is much larger than
the lower one. Hence, the upper component determines the FNS shift
of such a state. On the other hand, for a $p_{1/2}$ state the
lower component and hence its contribution to the FNS shift is
dominating. A straightforward calculation gives the following
values for the FNS shifts of $s$ and $p$ states (diagram
Fig.\ref{Fig1}(a))
\begin{eqnarray}
\label{E0}
 \Delta E_{ns_{1/2}}&=&\frac{2}{3n^3}(Z\alpha)^4m^3<r^2>\, ,\nonumber\\
 \Delta E_{np_{1/2}}&=&\frac{n^2-1}{6n^5}(Z\alpha)^6m^3<r^2>\, ,\nonumber\\
  \Delta E_{np_{3/2}}&=&\frac{n^2-1}{45n^5}(Z\alpha)^6m^5<r^4>\, .
\end{eqnarray}
Here $<r^2>$ and $<r^4>$ are values of $r^2$ and $r^4$   averaged
over charge density of the nucleus, $n$ is the principle quantum number.

The low-momentum expansion of the nuclear electric form
factor, see Eq. (\ref{dV1}), is of the form
\begin{equation}
\label{FF} {\cal F}(k^2)\approx
1-\frac{k^2}{6}<r^2>+\frac{k^4}{120}<r^4>\, .
\end{equation}
Modeling the nucleus as a uniformly charged ball  one gets
\begin{equation}
\label{r24} <r^2>= \frac{3}{5} \ r_0^2, \ \ \ <r^4>= \frac{3}{7} \ r_0^4 \ .
\end{equation}
As one should expect, the FNS corrections (\ref{E0}) obey the following
inequalities $\Delta E_{s_{1/2}}\gg\Delta E_{p_{1/2}}\gg\Delta E_{p_{3/2}}$.

We consider now radiative corrections which originate from quantum
fluctuations at distances $r \geq \lambda_C$. To calculate these "soft"
corrections it is sufficient to use nonrelativistic electron wave functions
and, instead of (\ref{dV}), to use an effective FNS perturbation
that reproduces the FNS correction $\Delta E_{ns1/2}$ for s-wave states,
see Eq. (\ref{E0}) and Ref.\cite{Mil2}. The effective perturbation reads
\begin{equation}
\label{g}
\delta V_{eff}(r)= g\delta(\bm r)\quad ,\quad g=\frac{2\pi
Z\alpha}{3} <r^2>\, .
\end{equation}
In the leading approximation, SEVFNS corrections for $2p_{1/2}$ and
$2p_{3/2}$ states, corresponding to diagrams
Fig.\ref{Fig1}(b,c), have been calculated in Refs.\cite{Mil2,Jen}:
\begin{eqnarray}
\label{Etot}
&&\delta E^{(0)}_{2p_{1/2}}=F\left[\ln\frac{1}{(Z\alpha)^2}+0.910\right]\,
 ,\nonumber \\
 &&\delta E^{(0)}_{2p_{3/2}}=F\left[\ln\frac{1}{(Z\alpha)^2}-0.215\right]\,
,\nonumber\\
&& where \ \  F=\frac{\alpha(Z\alpha)^5gm^3}{48\pi^2} \, .
\end{eqnarray}
These corrections are due to quantum fluctuations with the
frequency  $(Z\alpha)^2m \lesssim \omega \lesssim m$. Therefore,
the effective operator is calculated with the relativistic effects
taken into account. However, due to the behavior of the wave
functions of $p$-states at $r\ll \lambda_C/(Z\alpha)$, it is
possible to  average this operator in the leading approximation
over the  nonrelativistic wave function.
 For $s$-states the situation is quite different
\cite{LYE}.

 The ratio $\delta
E^{(0)}_{2p_{1/2}}/\Delta E_{2p_{1/2}}$ defined by
 Eqs. (\ref{Etot}) and (\ref{E0}) gives the relative correction (\ref{B0}).
This means that Eqs. (\ref{Etot}) correspond to zero order in
$(Z\alpha)$-expansion of the function $\mathcal{B}_{p_{1/2}}$

\begin{figure}[ht]
\centering
\includegraphics[height=180pt,keepaspectratio=true]{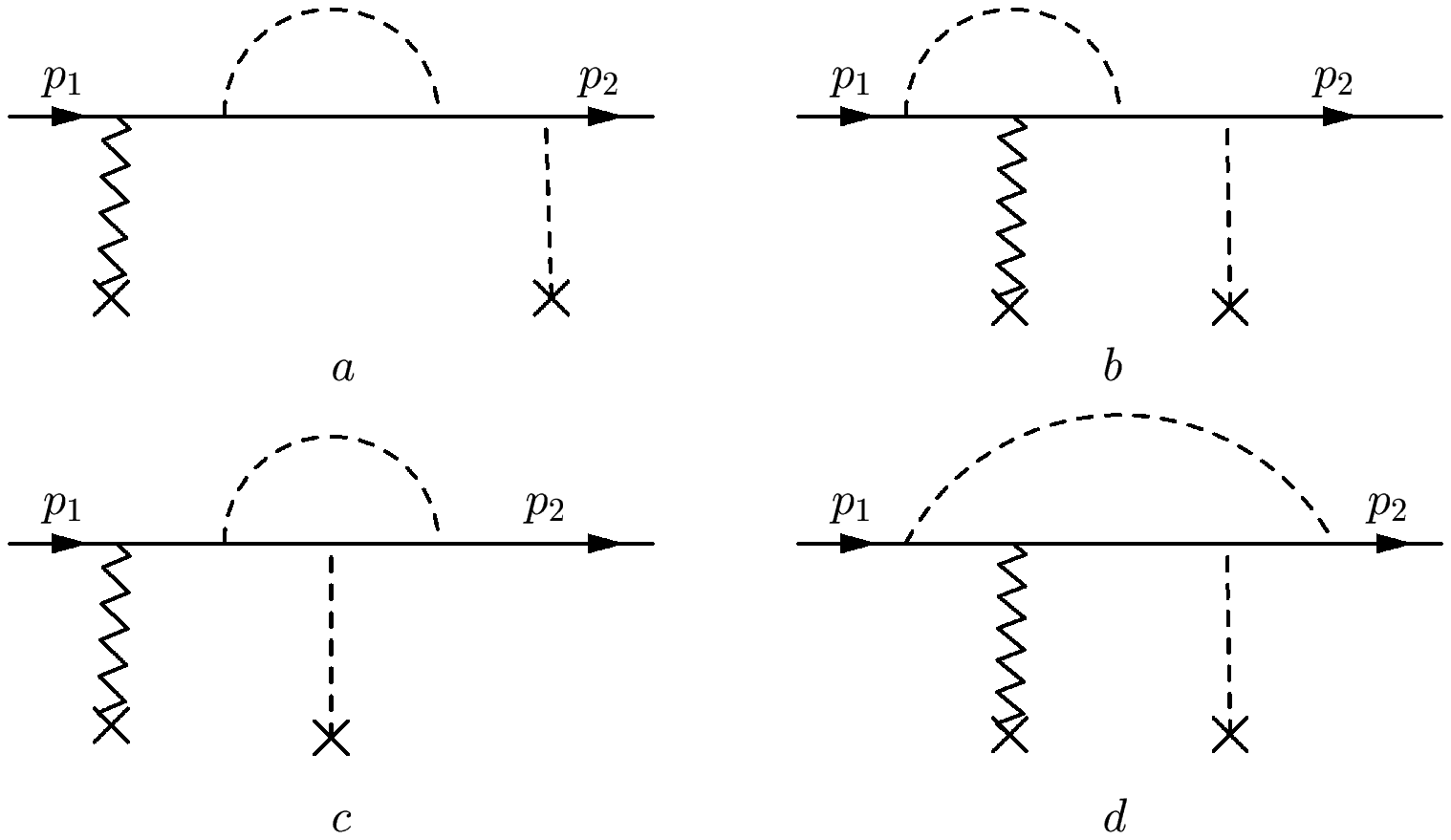}
\caption{\it Linear in $Z\alpha$ contributions to the electron
scattering amplitude related to the p-wave SEVFNS correction. The
solid line describes electron, the zigzag line denotes FNS
perturbation (\ref{g}), the dashed line denotes photon, and the
cross denotes the nucleus. The diagrams with the permutation of
Coulomb and perturbation lines should be added.} \label{diagrams}
\end{figure}

We will see that next-order corrections $\delta E^{(1)}_{p_{1/2}}$
and $\delta E^{(1)}_{p_{3/2}}$ do not contain any logarithms,
hence they come from quantum fluctuations at  $r\sim \lambda_C$.
Therefore it is convenient to calculate these corrections using
the effective operator approach, see, e. g. Ref. \cite{rev1}. In
this approach we first of all calculate  scattering amplitude,
described by diagrams shown in Fig.\ref{diagrams}.
External electron momenta in Fig.\ref{diagrams} are on mass shell
and it is sufficient to consider scattering at low momenta $|\bm
p_{1,2} |\ll m$. Here $\bm p_1 $ and $\bm p_2 $ are initial and
final momenta of the electron, respectively. To find the energy
shift one has to average the scattering amplitude over
nonrelataivistic wave function of atomic electron. We have already
used this technique to calculate (\ref{BB}) and to calculate a
similar term for atomic parity nonconservation effect, see Refs.
\cite{Mil,Mil1}. Similar to these papers we use here the
Fried-Yennie gauge \cite{Fried}. One can find expressions for
renormalized self-energy and vertex operators (ultraviolet
renormalization) in this gauge in Refs.\cite{rev1} and
\cite{Mil1}. However, the present calculation is more complicated
than that in  \cite{Mil,Mil1}. The point is that in Refs.
\cite{Mil,Mil1} it was sufficient to calculate only the forward
scattering amplitude, $\bm p_1= \bm p_2$. In the present case we
have to consider scattering at arbitrary angle, $\bm p_1 \ne \bm
p_2$. It is not just a simple technical complication. The forward
scattering amplitude is always infrared convergent. On the other
hand the finite angle amplitude is always infrared divergent
because of
 long range nature of the Coulomb field. To solve the problem we
regularize the Coulomb interaction in Fig.\ref{diagrams}, $1/{\bm
k}^2\rightarrow 1/({\bm k}^2+\lambda^2)$, where $m\gg \lambda\gg
|\bm p_{1,2}|$. Performing calculations we throw away all terms
inversely proportional to  $\lambda$ because they correspond not
to the true radiative corrections, but just to Coulomb  corrections to
the scattering amplitude corresponding to Eqs. (\ref{Etot}). Since we
perform  calculations in the Fried-Yennie gauge there is no need
in an infrared regularization of non-Coulomb photon propagator in
diagrams Fig.\ref{diagrams}. After a pretty long calculation we
have found the following expressions for diagrams in
Fig.\ref{diagrams}
\begin{eqnarray}
M_a&=&G\left[\frac{7}{16}\,(\bm\sigma\cdot
\bm p_2)(\bm\sigma \cdot\bm p_1)\right]\; , \nonumber\\
M_b+M_c&=&G\left[-\left(\frac{5}{12}+\frac{4}{3}\ln 2
\right)(\bm\sigma\cdot \bm p_2)(\bm\sigma \cdot\bm p_1)
+\left(\frac{227}{192}- \frac{1}{3}\ln2\right)(\bm
p_1\cdot\bm p_2) \right]\, ,\nonumber \\
M_d&=&G\left[\left(-\frac{1}{8}+\ln 2
\right)(\bm{\sigma}\cdot \bm{p}_2)(\bm{\sigma}
\cdot\bm{p}_1)-\left(\frac{61}{288}+
\frac{8}{3}\ln2\right)(\bm{p}_1\cdot\bm{p}_2) \right]\ .
\end{eqnarray}
Here $\bm\sigma$ is the Pauli matrix, and
$G=g\alpha(Z\alpha)/m^2$. Thus, the total scattering amplitude
corresponding to Fig.\ref{diagrams} reads
\begin{eqnarray}
\label{mtot}
&&M=M_a+M_b+M_c+M_d
= G\left[-\left(\frac{5}{48}+\frac{1}{3}\ln 2
\right)(\bm{\sigma}\cdot \bm{p}_2)(\bm{\sigma}
\cdot\bm{p}_1) +\left(\frac{559}{576}-
3\ln2\right)(\bm{p}_1\cdot\bm{p}_2) \right]\,.
\end{eqnarray}
To find the effective Hamiltonian we transfer (\ref{mtot}) to
coordinate space using substitutions $(\bm{\sigma}\cdot
\bm{p}_2)(\bm{\sigma} \cdot\bm{p}_1)\to  (\bm{\sigma}\cdot
\bm{p})\delta (\bm r)(\bm{\sigma} \cdot\bm{p}),$  and
$(\bm{p}_1\cdot\bm{p}_2)\to(\bm{p}\,\delta(\bm r)\bm{p}$) , where
$\bm p= -i \nabla$. Finally, calculating expectation values of the
effective Hamiltonian with the wave functions of $np$ states we
obtain the following corrections to energy levels :
\begin{eqnarray}
\label{Erad2}
 \delta E^{(1)}_{np_{1/2}}&=&F_1
\left(\frac{379}{432} - \frac{16}{3}\ln 2\right)\ ,\nonumber\\
 \delta E^{(1)}_{np_{3/2}}&=&F_1\left(\frac{559}{432}-
4\ln 2\right),\nonumber\\
F_1&=&\frac{\alpha(Z\alpha)^6gm^3(n^2\!-\!1)}
 {4\pi n^5}\, .
\end{eqnarray}
Using $ \delta E^{(0)}_{2p_{1/2}}$ from Eq. (\ref{Etot}),
$\delta E^{(1)}_{np_{1/2}}$ from Eq. (\ref{Erad2}),
$\Delta E_{np_{1/2}}$ from (\ref{E0}) as well as definitions
(\ref{ddef}), (\ref{general}) we find explicit expression
for $\mathcal{B}_{p1/2}$ valid up to the first order in $(Z\alpha)$.
\begin{eqnarray}
\label{B1}
\mathcal{B}_{p1/2}&=&-\alpha\left[-\frac{8}{9\pi}\left(\ln\frac{1}{(Z\alpha)^2}
+ 0.910\right)
+(Z\alpha)\left(-\frac{379}{432}+\frac{16}{3}\ln 2\right)\right]\ .
\end{eqnarray}
Strictly speaking the constant near the logarithm depends on the principal
quantum number. Equation (\ref{B1}) corresponds to the $2p_{1/2}$ state, for the
$3p_{1/2}$ state one has to replace $0.910 \to 0.908$, see Ref. \cite{Jen}.
Combining Eqs. (\ref{general}),(\ref{AA}) and(\ref{B1}) one gets the
the following expression for the SEVFNS radiative correction for a $p_{1/2}$
state
\begin{equation}
\label{psev}
\Delta^{SEV}_{p_{1/2}} = \alpha \frac{8}{9\pi}\left(\ln\frac{1}{(Z\alpha)^2}
+ 0.910\right)
-\alpha(Z\alpha)\left(-\frac{379}{432}+\frac{16}{3}\ln 2\right)
-\frac{\alpha(Z\alpha)^2}{\pi}\left(\frac{15}{4}-\frac{\pi^2}{6}\right)
\ln(b\lambda_C/r_0) \ .
\end{equation}
The leading unaccounted term is of the order $\sim
\alpha(Z\alpha)^2/\pi$. The correction (\ref{psev}) in \%  is
plotted in Fig.\ref{Fig3} by the solid line. The same correction,
but with $\mathcal{A}$ taken from Fig.\ref{Fig2} is plotted in
Fig.\ref{Fig3} by the dashed-dotted line. Our analytical
calculation of $\Delta_{p_{1/2}}^{SEV}$  agrees very well with
previous numerical results for $2p_{1/2}$ states shown by diamonds
\cite{CJS93} and crosses \cite{Mohr}, see also comment
\cite{com1}.

Let us discuss now the SEVFNS correction for a $p_{3/2}$ state. In
this case the definition (\ref{ddef}) for the relative value of
the correction is not sensible. The point is that according to
(\ref{Etot}) and (\ref{Erad2}) the SEVFNS correction $\delta
E_{p_{3/2}} \propto \langle r^2 \rangle$, and according to
(\ref{E0}) the FNS correction $\Delta E_{p_{3/2}} \propto \langle
r^4 \rangle$. Therefore the definition (\ref{ddef}) would imply
that $\Delta_{p_{3/2}} \gg 1$. This itself is not a problem, just
some rescaling. The real problem is that different powers of $r^2$
indicate that relativistic factors for $\delta E_{p_{3/2}}$ and
$\Delta E_{p_{3/2}}$ are different. Therefore convergence of
$(Z\alpha)$-expansion for $\Delta_{p_{3/2}}$ defined according to
Eq. (\ref{ddef}) must be very poor. Instead we define
$\Delta_{p_{3/2}}$ as
\begin{equation}
\label{ddef1}
\Delta_{np_{3/2}}=\delta E_{np_{3/2}}/\Delta E_{np_{1/2}} \ .
\end{equation}
This is the ratio of the SEVFNS energy shift for $np_{3/2}$ state
and the FNS energy shift for $np_{1/2}$ state. Using definition
(\ref{ddef1}) we absorb all small-distance ($r \sim r_0$) effects
into $\Delta E_{np_{1/2}}$, and therefore, according to the
effective operator approach explained in section II, we expect a
reasonable convergence of $(Z\alpha)$-expansion for
$\Delta_{np_{3/2}}$. Using $ \delta E^{(0)}_{2p_{3/2}}$ from Eq.
(\ref{Etot}), $\delta E^{(1)}_{np_{3/2}}$ from Eq. (\ref{Erad2}),
$\Delta E_{np_{1/2}}$ from (\ref{E0}) as well as the definition
(\ref{ddef1}) we obtain
\begin{eqnarray}
\label{B13}
\Delta_{p_{3/2}}^{SEV}&=&\alpha\frac{8}{9\pi}\left(\ln\frac{1}{(Z\alpha)^2}
-0.215\right)-\alpha(Z\alpha)\left(-\frac{559}{432}+4\ln 2\right)\ .
\end{eqnarray}
Again, the constant near the logarithm depends on the principal
quantum number. Equation (\ref{B13}) corresponds to the $2p_{3/2}$
state, for the $3p_{3/2}$ state one has to replace $-0.215 \to
-0.217$, see Ref. \cite{Jen}. There are no $\ln(\lambda_C/r_0)$
terms in $\Delta_{p_{3/2}}$, so the leading unaccounted term in
(\ref{B13}) is of the order $\sim \alpha(Z\alpha)^2/\pi$. The
correction $\Delta_{p_{3/2}}^{SEV}$ (\%) given by Eq. (\ref{B13})
is plotted in Fig.\ref{Fig3} by the long-dashed line. Note that it
is comparable to $\Delta_{p_{1/2}}^{SEV}$. It means that absolute
SEVFNS energy shifts $\delta E_{p_{3/2}}$ and $\delta E_{p_{1/2}}$
are also comparable.

\section{Vacuum polarization FNS radiative corrections}\label{IV}
\begin{figure}[ht]
\centering
\includegraphics[height=100pt,keepaspectratio=true]{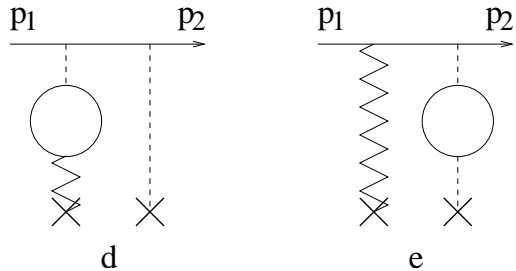}
\caption{\it Linear in $Z\alpha$ FNS vacuum
polarization corrections corresponding to $Z\alpha$-expansion of
diagrams Fig.\ref{Fig1}d and Fig.\ref{Fig1}e. The solid line describes
electron, the zigzag line denotes FNS perturbation (\ref{g}), the dashed line
denotes photon, and the cross denotes the nucleus.}
\label{FZA}
\end{figure}

In this section we consider vacuum polarization FNS  radiative corrections
(VPFNS) given by diagrams in Fig.\ref{Fig1}d and Fig.\ref{Fig1}e.
Let us start from $s_{1/2}$ states. In the leading order in $(Z\alpha)$
the diagrams Fig.\ref{Fig1}d and Fig.\ref{Fig1}e can be represented as it
is shown in Fig.\ref{FZA}.
Calculation of these diagrams for forward scattering $\bm p_1=\bm p_2$ is very
simple, they give equal contributions, and their total contribution to the s-wave
VPFNS correction is \cite{F,H}
\begin{equation}
\label{FH}
\Delta^{VP(1)}_{s_{1/2}}=\frac{3}{4}\alpha(Z\alpha) \ .
\end{equation}
Logarithmically enhanced contribution  of diagram Fig.\ref{Fig1}e
has been calculated in Ref.\cite{Mil0}, it reads
\begin{equation}
\label{elog}
\Delta^{(e,log)}=\frac{2\alpha(Z\alpha)^2}{3\pi\gamma}\ln^2
\left(\frac{b\lambda_C}{r_0}\right)\  .
\end{equation}
This formula is equally valid for $s_{1/2}$ and $p_{1/2}$ states,
for definition of parameters  see Eq.(\ref{general}).
A logarithmically enhanced contribution of the diagram Fig.\ref{Fig1}d
has not been calculated yet. A similar contribution has been estimated
in Refs.\cite{Mil,Mil1} in relation to atomic parity nonconservation (PNC).
For PNC this contribution is strongly suppressed by the factor
$1-4\sin^2\theta_W$, where $\theta_W$ is the Weinberg angle, therefore
there have been no need in an accurate calculation. Now we perform
an accurate  calculation, the calculation is equally valid for $s_{1/2}$ and $p_{1/2}$
states.

According to (\ref{Dirac}) the electron density at $r\ll
(Z\alpha)\lambda_C$ is
\begin{equation}
\label{rr}
\rho(r)=M r^{2(\gamma-1)} \ ,
\end{equation}
where $M$ is some constant.
Fourier transform of the density reads
\begin{eqnarray}
\label{rk}
\int e^{i\bm k\bm r}\, \rho(r)d\bm r=
M\frac{4\pi}{k^{2\gamma+1}}\sin{(\pi\gamma)}\, \Gamma(2\gamma) \ ,
\end{eqnarray}
where $\Gamma(x)$ is the gamma function. Then, due to Eqs.
(\ref{dV}) and (\ref{dV1}) the FNS energy shift given by diagram
Fig.\ref{Fig1}a reads
\begin{eqnarray}
\label{des} \Delta E=\int \delta V(r) \rho(r) d\bm r= M
4Z\alpha\sin{(\pi\gamma)}\Gamma(2\gamma) r_0^{2\gamma}
\int_0^\infty \frac{dy}{y^{\gamma+1}}[1-{\cal F}_1(y)]\, .
\end{eqnarray}
where $y=k^2r_0^2$, and ${\cal F}_1(y)={\cal F}(y/r_0^2)$.

To generate diagram Fig.\ref{Fig1}d one has to account for the
electron loop in the interaction energy (\ref{des}). To do so we
perform the following replacement in (\ref{des}), see, e.g.,
Ref. \cite{rev1} and comment \cite{com2}
\begin{eqnarray}
\label{repl}
\frac{1}{k^2} \to \frac{\alpha}{\pi m^2}\int_0^1dv\frac{v^2(1-v^2/3)}{4+(1-v^2)k^2/m^2} \ .
\end{eqnarray}
This leads to the following expression for diagram Fig.\ref{Fig1}d
\begin{eqnarray}
\label{ded} \delta E^{(d)}=M
4Z\alpha\sin{(\pi\gamma)}\Gamma(2\gamma) r_0^{2\gamma}
\int_0^\infty \frac{dy}{y^{\gamma+1}}[1-{\cal F}_1(y)]\,\,
\frac{\alpha}{\pi}\int_0^1 \frac{dv\,
v^2(1-v^2/3)(y/a^2)}{4+(1-v^2)(y/a^2)}\, ,
\end{eqnarray}
where $a=mr_0=r_0/\lambda_C \ll 1$. Having in mind this inequality the expression
(\ref{ded}) can be transformed to
\begin{eqnarray}
\label{ded1}
\delta E^{(d)}&=&M 4Z\alpha\sin{(\pi\gamma)}\Gamma(2\gamma) r_0^{2\gamma}
\left(\frac{\alpha}{3\pi}\right)\left\{\int_0^\infty
\frac{dy}{y^{\gamma+1}}[1-{\cal F}_1(y)]\left[\ln\frac{y}{a^2}-
\frac{5}{3}\right]\right.\nonumber\\
&+&\left.\frac{3{\cal F}_1'(0)\, a^{2(1-\gamma)}\,\pi^{3/2}
4^{1-\gamma}\,\Gamma(\gamma+1)}{4(\gamma-1)\,\sin(\pi\gamma)
\,\Gamma(\gamma+3/2)}\right\}\, ,
\end{eqnarray}
where ${\cal F}_1'(0) =\left.\frac{d{\cal F}_1}{dy}\right|_{y=0}$.
Finally, introducing notation
\begin{eqnarray}
\label{j} J=\int_0^\infty\frac{dy}{y^{\gamma+1}}[1-{\cal F}_1(y)]
\, ,
\end{eqnarray}
and substituting Eqs. (\ref{ded1}) and (\ref{des}) in the definition
(\ref{ddef}) we obtain the following expression for logarithmically
enhanced contribution to the relative correction corresponding to Fig.\ref{Fig1}d
\begin{eqnarray}
\label{VPd}
\Delta^{(d,log)}=\left(\frac{\alpha}{3\pi}\right)\left\{
\left[2\ln\frac{\lambda_C}{r_0}-\frac{5}{3}-\frac{\partial\, \ln
J} {\partial\gamma }\right] +\frac{3{\cal
F}_1'(0)(\lambda_C/r_0)^{2(1-\gamma)}\,\pi^{3/2}
4^{1-\gamma}\,\Gamma(\gamma+1)}{4(\gamma-1)\,\sin(\pi\gamma)
\,\Gamma(\gamma+3/2)\,J}\right\}\, .
\end{eqnarray}
This formula is equally valid for $s_{1/2}$ and $p_{1/2}$ states.

For large $Z$, as soon as $(1-\gamma)\ln(\lambda_C/r_0) \gg 1$,
the ${\cal F}_1'$-term in (\ref{VPd}) can be neglected and
\begin{eqnarray}
\label{VPd1}
\Delta^{(d,log)}\to\left(\frac{\alpha}{3\pi}\right)
\left[2\ln\frac{\lambda_C}{r_0}-\frac{5}{3}-\frac{\partial\, \ln
    J}{\partial\gamma }\right] \ .
\end{eqnarray}
A similar high $Z$ estimate has been used in \cite{Mil,Mil1}. For small $Z$,
more precisely for $(1-\gamma)\ln(\lambda_C/r_0) \ll 1$, the correction
behaves as
\begin{eqnarray}
\label{VPd2}
\Delta^{(d,log)}\to \frac{\alpha(Z\alpha)^2}{3\pi} \ln^2\left(\frac{\lambda_C}
{r_0}\right)\, .
\end{eqnarray}
The simplest way to obtain this limiting behavior is to expand intermediate expressions
(\ref{des}) and(\ref{ded}), however, one can also obtain the same result
expanding the exact answer (\ref{VPd}).

\begin{figure}[ht]
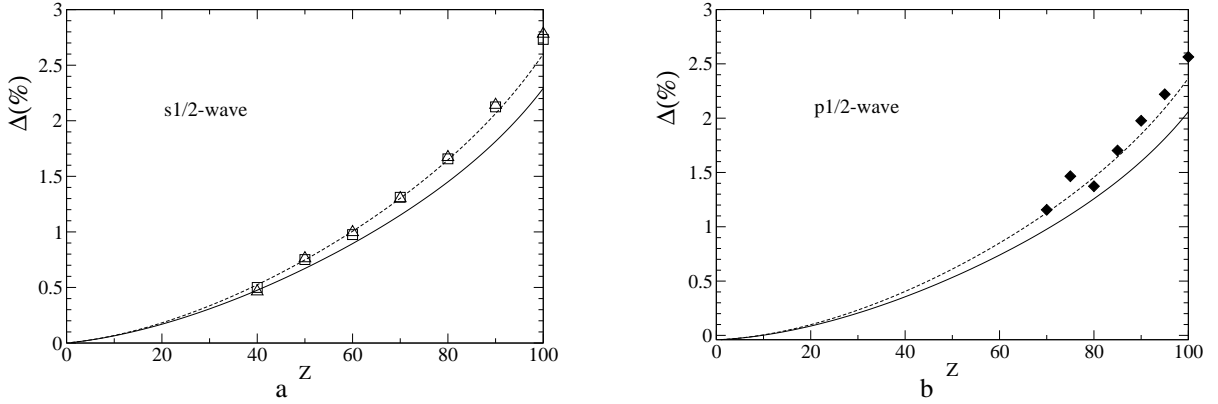

\centering
\vspace{15pt}
\includegraphics[height=150pt,keepaspectratio=true]{Fig6s.eps}
\hspace{30pt}
\includegraphics[height=150pt,keepaspectratio=true]{Fig6p.eps}
\caption{\it Relative VPFNS corrections (\%) for
$s_{1/2}$ and $p_{1/2}$ states. The solid line in Fig. ``a''
shows the correction $\Delta_s^{VP}$ given by Eq. (\ref{VP}) with $u(Z\alpha)=0$.
The dashed line shows the same correction with $u(Z\alpha)$ given by
Eq. (\ref{u}).
Results of computations \cite{JS}  are shown by
squares for $1s$ states and by triangles for $2s$ states.
The solid line in Fig. ``b'' shows the correction
$\Delta_{p_{1/2}}^{VP}$ given by Eq. (\ref{VPp})  with $u(Z\alpha)=0$.
The dashed line shows the same correction with $u(Z\alpha)$ given by Eq. (\ref{u}).
Results of computations \cite{JS} for $2p_{1/2}$ states are shown by diamonds.
} \label{Fig6}
\end{figure}

Behavior of the exact expression (\ref{VPd}) between asymptotical
regimes (\ref{VPd1}) and (\ref{VPd2}) depends on the particular
form factor $F$. However, the dependence is pretty weak, and for
further analysis we use the form factor of uniformly charged ball
of radius $r_0$.
\begin{eqnarray}
\label{fb}
F(y)=\frac{3}{y}\left[\frac{\sin\sqrt{y}}{\sqrt{y}}-\cos\sqrt{y}\right]
\end{eqnarray}
An explicit integration in (\ref{j}) gives
\begin{eqnarray}
\label{jj}
&&J=\frac{6\cos(\pi\gamma)\Gamma(-1-2\gamma)}{3+2\gamma}\, ,
\nonumber\\
&&-\frac{\partial \ln J}{\partial \gamma}=\pi \tan(\pi\gamma)+
\frac{2}{3+2\gamma}+2\psi(-1-2\gamma)\, ,
\end{eqnarray}
where $\psi(x)=d\ln\Gamma(x)/d x$. Combining together Eqs.
(\ref{FH}), (\ref{elog}),  (\ref{VPd}), (\ref{jj}), as well as
unaccounted in the present calculation correction $u(Z\alpha)\sim
\alpha(Z\alpha)^2/\pi$ we obtain the following expression for the
total vacuum polarization correction for s-electron
\begin{eqnarray}
\label{VP}
\Delta^{VP}_{s_1/2}&=&\frac{3}{4}\alpha(Z\alpha)
+\frac{2\alpha(Z\alpha)^2}{3\pi\gamma}\ln^2
\left(\frac{b\lambda_C}{r_0}\right)\nonumber\\
&+&\left(\frac{\alpha}{3\pi}\right)\left\{
\left[2\ln\frac{\lambda_C}{r_0}-\frac{5}{3}+\pi \tan(\pi\gamma)+
\frac{2}{3+2\gamma}+2\psi(-1-2\gamma)\right]\right.\nonumber\\
&-&\left.
\frac{\pi^{3/2}4^{(1-\gamma)}(3+2\gamma)\Gamma(\gamma+1)}
{40\sin(2\pi\gamma)(\gamma-1)\Gamma(-1-2\gamma)\Gamma(\gamma+3/2)}
\left(\frac{\lambda_C}{r_0}\right)^{2(\gamma-1)}\right\} +
u(Z\alpha)\, .
\end{eqnarray}
The correction (\ref{VP}) with $u(Z\alpha)=0$ is plotted in Fig.\ref{Fig6}a by the solid line.
Results of computations of the same correction performed in Ref \cite{JS}
for $1s$ states are shown by  squares and for $2s$ states are shown by
triangles. Agreement between our analytical calculation and \cite{JS})
is pretty good. Moreover, we can determine $u(Z\alpha)$
fitting the numerical data. Result of the fit reads
\begin{equation}
\label{u}
u=2.5\frac{\alpha(Z\alpha)^2}{\pi} \ .
\end{equation}
The correction (\ref{VP}) with $u(Z\alpha)$ given by (\ref{u}) is plotted in
Fig.\ref{Fig6}a by the dashed line.

Let us consider now the VPFNS correction for a $p_{1/2}$ state.
Logarithmically enhanced contributions (\ref{elog}) and (\ref{VPd}) are
exactly the same as for $s_{1/2}$ states. There is also a contribution
of zero order in $(Z\alpha)$ calculated in \cite{Mil2}
\begin{equation}
\label{vp0}
\Delta^{VP(0)}_{p_{1/2}}=-\frac{8}{45\pi}\alpha \ .
\end{equation}
To calculate the first order correction we once more employ the
effective operator approach and consider the scattering amplitude
described by Fig.\ref{FZA}, in this case ${\bm p}_1 \ne {\bm
p}_2$. Similar to the procedure described in Section III we
regularize the Coulomb interaction in Fig.\ref{FZA}, $1/{\bm
k}^2\rightarrow 1/({\bm k}^2+\lambda^2)$, where $m\gg \lambda\gg
|\bm p_{1,2}|$. Performing calculations we throw away all terms
inversely proportional to  $\lambda$ because they correspond not
to true radiative corrections, but just to Coulomb  corrections to
scattering amplitude corresponding to Eq. (\ref{vp0}). Then we
obtain the following result for the first-order correction
\begin{equation}
\label{vp1}
\Delta^{VP(1)}_{p_{1/2}}=\frac{23}{72}\alpha(Z\alpha) \ .
\end{equation}
The coefficient is $23/72=1/8+7/36$, where the first term comes from the
diagram Fig.\ref{FZA}e, and the second term comes from the diagram Fig.\ref{FZA}d.
Combining together Eqs. (\ref{vp0}), (\ref{vp1}), (\ref{elog}),  (\ref{VPd}), (\ref{jj}),
as well as unaccounted in the present calculation correction
 $u(Z\alpha)\sim \alpha(Z\alpha)^2/\pi$ we obtain the following expression for
 total  vacuum   polarization correction for a $p_{1/2}$ state
\begin{eqnarray}
\label{VPp}
\Delta^{VP}_{p_{1/2}}&=&-\frac{8}{45\pi}\alpha +\frac{23}{72}\alpha(Z\alpha)
+\frac{2\alpha(Z\alpha)^2}{3\pi\gamma}\ln^2
\left(\frac{b\lambda_C}{r_0}\right)\nonumber\\
&+&\left(\frac{\alpha}{3\pi}\right)\left\{
\left[2\ln\frac{\lambda_C}{r_0}-\frac{5}{3}+\pi \tan(\pi\gamma)+
\frac{2}{3+2\gamma}+2\psi(-1-2\gamma)\right]\right.\nonumber\\
&-&\left.
\frac{\pi^{3/2}4^{(1-\gamma)}(3+2\gamma)\Gamma(\gamma+1)}
{40\sin(2\pi\gamma)(\gamma-1)\Gamma(-1-2\gamma)\Gamma(\gamma+3/2)}
\left(\frac{\lambda_C}{r_0}\right)^{2(\gamma-1)}\right\} +
u(Z\alpha)\, .
\end{eqnarray}
In principle the unaccounted contribution $u(Z\alpha)$ can be
different from that in (\ref{VP}). The correction (\ref{VPp}) with
$u(Z\alpha)=0$ is plotted in Fig.\ref{Fig6}b by a solid line.
Results of computations of $\Delta^{VP}_{p_{1/2}}$ performed in
Ref \cite{JS} for $2p_{1/2}$ states are shown by  diamonds. Once
more, the agreement is pretty good. We also find unknown
$u(Z\alpha)$ fitting the numerical data \cite{JS}. The fit gives
the result very close to that for s-wave, see Eq. (\ref{u}). The
dashed line in Fig.\ref{Fig6}b shows the correction (\ref{VPp})
with $u(Z\alpha)$ given by Eq. (\ref{u}).

Finally we consider the VPFNS correction for a $p_{3/2}$ state, the relative correction
is defined according to Eq. (\ref{ddef1}). There is a contribution of zero order in
$(Z\alpha)$ calculated in \cite{Mil2}
\begin{equation}
\label{vp03}
\Delta^{VP(0)}_{p_{3/2}}=-\frac{8}{45\pi}\alpha \ .
\end{equation}
It is equal to that for a $p_{1/2}$ state, see Eq. (\ref{vp0}).
The first-order correction is given by diagrams in  Fig.\ref{FZA}.
The calculation of the correction, which is similar to the
calculation of (\ref{vp1}), gives
\begin{equation}
\label{vp13} \Delta^{VP(1)}_{p_{3/2}}=\frac{5}{72}\alpha(Z\alpha) \ .
\end{equation}
This  correction  comes from the diagram Fig.\ref{FZA}d only.
There are no logarithmically enhanced contributions to $\Delta^{VP}_{p_{3/2}}$,
therefore with accuracy $\sim \alpha(Z\alpha)^2/\pi$ accepted in the present work
the correction $\Delta^{VP}_{p_{3/2}}$ is given by sum of (\ref{vp03}) and (\ref{vp13})
\begin{equation}
\label{vp3}
\Delta^{VP}_{p_{3/2}}=-\frac{8}{45\pi}\alpha+\frac{5}{72}\alpha(Z\alpha)
\ .
\end{equation}

\section{Summary and conclusions}\label{V}
In the present work we have calculated analytically the finite nuclear
size effect on Lamb shift of $s_{1/2}$, $p_{1/2}$, and $p_{3/2}$
atomic states. Corresponding radiative corrections are strongly
affected by ultrarelativistic behavior of electrons in the
vicinity of the nucleus. As a result of this behavior the convergence
of $Z\alpha$-expansion is very poor for  absolute energy shifts,
and on the other hand the convergence is reasonably good for
relative corrections which we consider. We define the relative
corrections for $s_{1/2}$ and $p_{1/2}$ states according to Eq.
(\ref{ddef}), and for $p_{3/2}$ states according to Eq.
(\ref{ddef1}). The self-energy and vertex finite size (SEVFNS)
relative radiative correction  for a $s_{1/2}$ state is given by
Eq. (\ref{ssev}). The SEVFNS correction for a $p_{1/2}$ state is
given by Eq. (\ref{psev}),  and the SEVFNS correction for a
$p_{3/2}$ state is given by Eq. (\ref{B13}). The vacuum
polarization finite size (VPFNS) relative radiative correction for
a $s_{1/2}$ state is given by Eqs. (\ref{VP}),(\ref{u}). The VPFNS
correction for a $p_{1/2}$ state is given by Eqs.
(\ref{VPp}),(\ref{u}),  and the VPFNS correction for a $p_{3/2}$
is given by Eq. (\ref{vp3}). Our analytical results for $s_{1/2}$
and $p_{1/2}$ states agree well with previous numerical
calculations.

\begin{figure}[ht]
\centering
\vspace{20pt}
\includegraphics[height=200pt,keepaspectratio=true]{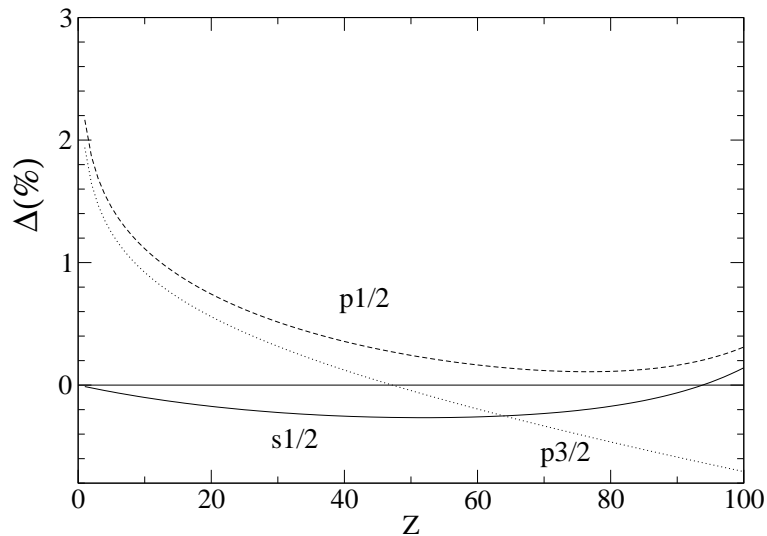}
\vspace{10pt}
\caption{\it Total (SEVFNS + VPFNS) finite size relative
radiative corrections (\%) for $s_{1/2}$ (solid line),
$p_{1/2}$ (dashed line), and $p_{3/2}$ (dotted line) states.
} \label{Fig7}
\end{figure}

Finally, in Fig. \ref{Fig7} we plot the total radiative
corrections which include self-energy, vertex and vacuum
polarization diagrams (SEVFNS + VPFNS). For $s_{1/2}$ and
$p_{1/2}$ states at large $Z$ there are strong compensations
between SEVFNS and VPFNS contributions. We would like to attract
attention to the $p_{3/2}$ correction. The {\it same}
normalization is used for $p_{1/2}$ and $p_{3/2}$ states.
Therefore, from Fig. \ref{Fig7}  we come to a somewhat paradoxical
conclusion that at large $Z$ the finite nuclear size effect on
Lamb shift of $p_{3/2}$ states is larger than that for $p_{1/2}$
states. Moreover, using Fig. \ref{Fig7} and Eqs. (\ref{Sh}) we
conclude that at large $Z$ the finite nuclear size effect on Lamb
shift of $p_{3/2}$ states is only by a few times smaller than that
for $s$ states. These conclusions are not consistent with
numerical data \cite{JS} at large $Z$. Certainly, our calculation
for $p_{3/2}$ states is valid only up to the order
$\alpha(Z\alpha)$. The $p_{3/2}$ correction does not contain the
ultraviolet logarithm and therefore the convergence of
$Z\alpha$-expansion can be different from that for $s_{1/2}$ and
$p_{1/2}$ states. A poor convergence of the expansion could be the
reason for disagreement with previous numerical calculatins.
However a new exact in $Z\alpha$ numerical calculation for
$p_{3/2}$ states would be
interesting.\\

We are grateful to G. Soff for helpful comments.
 The work was supported by RFBR grants 01-02-16926 and 03-02-16510.


\begin{thebibliography}{99}
\bibitem{com0} Except of diagrams shown in Fig. \ref{Fig1} there is
also a contribution related to normalization of the bound state wave
function (a "derivative term").  This term is not important for
calculations performed in the present work.
However, it is important for calculations performed in our
previous papers \cite{Mil,Mil1} results of which we use here.
The ``derivative term'' has been taken into account in
\cite{Mil,Mil1}.
\bibitem{com1} In the previous work \cite{Mil2} we found the
$\alpha(Z\alpha)$-term in Eqs. (\ref{B1}),(\ref{psev}) by
fitting numerical data \cite{CJS93,Mohr}. As a result of the fit we got
$-2.75\alpha(Z\alpha)$. According to the calculation performed
in the present work the exact result reads
$(379/432-16/3\ln 2)\alpha(Z\alpha)=-2.819\alpha(Z\alpha)$.
\bibitem{com2} In this work we do not consider Wichmann-Kroll
corrections. These corrections to the induced-charge density
related to the FNS effect were considered in \cite{Lee}.
\end{thebibliography}
\end{document}